\documentclass[rnote]{aa} % for the research notes
\usepackage{graphicx}%,textcomp}
\usepackage{txfonts}
\newcommand{\be}{\,\begin{equation}}
\newcommand{\ee}{\,\end{equation}}

\begin{document} 

   \title{Non-linear cosmic ray Galactic transport\\ in the light of AMS-02 and Voyager data}

%   \subtitle{I. Overviewing the $\kappa$-mechanism}

   \author{R. Aloisio,
          \inst{1,2}
  	P. Blasi
	\inst{2,1}
          \and
          P. D. Serpico \inst{3}
          }

   \institute{Gran Sasso Science Institute (INFN), Viale F. Crispi 7, 67100 L\textquoteright%\textquotesingle 
   Aquila, Italy\\
              \email{roberto.aloisio@gssi.infn.it}
         \and
          INAF/Osservatorio Astrofisico di Arcetri, Largo E. Fermi, 5 50125 Firenze, Italy \\
             \email{blasi@arcetri.astro.it}
	\and
	LAPTh, Univ. Savoie Mont Blanc, CNRS, B.P.110, Annecy-le-Vieux F-74941, France\\
	\email{serpico@lapth.cnrs.fr}
             }

%   \date{Received:; accepted:}

% \abstract{}{}{}{}{} 
% 5 {} token are mandatory
 
  \abstract
   {Features in the spectra of primary cosmic rays (CRs) provide invaluable information on the propagation of these particles in the Galaxy. In the rigidity region around a few hundred GV, such features have been measured in the proton and helium spectra by the PAMELA experiment and later confirmed with a higher significance by AMS-02. We investigate the implications of these datasets for the scenario in which CRs propagate under the action of self-generated waves.}
   {We show that the recent data on the spectrum of protons and helium nuclei as collected with AMS-02 and Voyager are in very good agreement with the predictions of a model in which the transport of Galactic CRs is regulated by self-generated waves. We also study the implications of the scenario for the boron-to-carbon ratio: although a good overall agreement is found, at high energy we find marginal support for a (quasi) energy independent contribution to the grammage, that we argue may come from the sources themselves.}
   {The transport equation for both primary and secondary nuclei is solved together with an equation for the evolution of the self-generated waves and a background of pre-existing waves. The solution of this system of non-linear equations is found with an iterative method elaborated by the same authors in previous work on this topic.}
   {A break in the spectra of all nuclei is found at rigidity of a few hundred GV, as a result of a transition from self-generated waves to pre-existing waves with a Kolmogorov power spectrum. Neither the slope of the diffusion coefficient, nor its normalisation are free parameters. Moreover, at rigidities below a few GV, CRs are predicted to be advected with the self-generated waves at the local Alfv\'en speed. This effect, predicted in our previous work, provides an excellent fit to the Voyager data on the proton and helium spectra at low energies, providing additional support to the model.}
   {}


\date{Received; accepted\\
Preprint numbers: LAPTH-036/15}
\maketitle
%
%________________________________________________________________

\section{Introduction}
\label{sec:intro}

The transport of Galactic cosmic rays (CRs) is likely to be very complex: the structure of the large scale magnetic field is complicated and very poorly known, and the structures on small scales (resonant with the CR particles) are basically unknown, although the fact that the diffusion paradigm seems to work can be considered as an indirect evidence for the existence of such small scale turbulence. The origin of the power on small scales is also unknown, but one contribution that seems to be hardly avoidable is that due to the self-generation of perturbations due to the CR current, proportional to the gradient of CRs, which in turn is due to the existence of the same scattering centres, responsible for diffusion. This simple description  is sufficient to emphasise the non-linear nature of this process, which is qualitatively similar to what happens at supernova shocks, thought to be the main sources of Galactic CRs (see the recent review by \cite{blasirev}). In addition to the self-generated waves,  turbulence at larger spatial scales is generically expected and becomes important for scattering CRs of higher energies. 

This scenario has been analysed in detail in \cite[]{serpico1,aloisio1}, where the main implications were discussed: \cite{serpico1} calculated the spectrum of protons under the action of both the self-generated and pre-existing turbulence, and compared the results with the PAMELA data available at the time \cite[]{PAMELAbreak}, where the first direct detection of a spectral break in proton and helium fluxes at few hundred GV was claimed. The first release of the data collected by AMS-02 \cite[]{AMS13p,AMS13He} did not confirm the existence of these spectral features and brought the investigation on this topic to an almost complete standstill, waiting for the resolution of the observational conundrum. Recently the AMS collaboration published the final analysis of the data on the proton spectrum \cite[]{ams02}, where a change of slope at few hundred GV is evident. At the present time, only preliminary results on the spectrum of helium and carbon nuclei are available \cite[]{amsslides}, but a similar break is visible in such data as well. Moreover, preliminary data on the B/C ratio have also been presented: the small statistical error bars up to high energies allow us to use this tool as a powerful indicator of the propagation of CRs through the Galaxy. 

In addition to the AMS-02 data, the results of another invaluable experimental effort became available in the last few years: the Voyager spacecraft, launched in 1977, reached the termination shock and is now believed to be moving in the interstellar medium, unaffected by the solar wind \cite[]{voyager}. The Voyager is therefore providing us with the very first measurement of the interstellar spectra of protons and helium nuclei, that can be compared directly with the predictions of our models, rather than dealing with complex and uncertain recipes of solar modulation. 

These recent developments revived the interest of the Community and stimulated a new search for explanations of the spectral breaks: in this paper we reconsider the model first put forward by \cite{serpico1,aloisio1} and check it versus both the data of AMS-02 and Voyager. We find that the set of parameters that were previously used to fit the PAMELA data also lead to a good fit to the AMS-02 data on the proton and helium spectra. The rigidity where the spectral break occurs is predicted rather solidly by the model. In fact the model predicts in a rather constrained manner both the normalisation and the energy dependence of the diffusion coefficient at rigidities below a few hundred GV, where self-generation is important. At the same time it also fixes the injection rate (for instance in SNR shocks) and hence the diffusion coefficient at high energies, in order to reproduce the observed spectrum of CRs. Since both contributions are quite well constrained, so is the transition energy. The constrained nature of the model is also its main strength. As discussed above, the B/C ratio was presented by the AMS-02 collaboration only in a preliminary form~\cite[]{amsslides} and we use this information only to further check, at least visually, the viability of the model: the AMS data extend to unprecedentedly high energies and hence they provide us with much information about propagation. We find that the preliminary data on B/C are well fit with our model, although there is marginal evidence for some additional grammage with a harder energy dependence compared with the grammage accumulated throughout the Galaxy. The necessary grammage is compatible with that accumulated by CRs while being accelerated and advected in a typical SNR, and is consistent with being energy independent. 

The paper is structured as follows: in \S \ref{sec:calculations} we summarize the essential aspects of the model put forward by \cite{serpico1,aloisio1}. In \S \ref{sec:results} we discuss the implications for the nuclear spectra and for the B/C ratio. We summarize in \S \ref{sec:summary}.

\section{Self-Generated versus pre-existing waves and CR transport}
\label{sec:calculations}

In a 1D model (infinite slab, only dependence from distance $z$ above/below the plane retained), the transport equation for CR nuclei in its general form can be written as follows:
$$
-\frac{\partial}{\partial z} \left[D_{\alpha}(p) \frac{\partial f_{\alpha}}{\partial z}\right] + v_{A} \frac{\partial f_{\alpha}}{\partial z}-\frac{2}{3}v_{A}\delta(z) p \frac{\partial f_{\alpha}}{\partial p}
+\frac{\mu v(p) \sigma_{\alpha}}{m}\delta(z) f_{\alpha} + 
$$
$$
\frac{1}{p^{2}} \frac{\partial}{\partial p}\left[ p^{2} \left(\frac{dp}{dt}\right)_{\alpha,ion} f_{\alpha}\right] =
$$
\be
~~~~~~~~~~~~= 2 h_d q_{0,\alpha}(p) \delta(z) +\sum_{\alpha'>\alpha} \frac{\mu\, v(p) \sigma_{\alpha'\to\alpha}}{m}\delta(z) f_{\alpha'},
\label{eq:slab}
\ee
where $\sigma_{\alpha}$ is the spallation cross section of a nucleus of type $\alpha$, $\mu$ is a grammage parameter fixed to 2.4 mg/cm$^2$, and $q_{0,\alpha}(p)$ is the rate of injection per unit volume in the disc of the Galaxy. Namely,
since $h_d$ is the half-thickness of the (assumed infinitesimal) gaseous disk, $2\,h_d\,q_{0,\alpha}$ is the rate of injection in the disc of the Galaxy per unit surface.
The total cross section for spallation and the cross sections for the individual channels of spallation of a heavier element to a lighter element ($\sigma_{\alpha'\to\alpha}$) have been taken from \cite{Webber:1990p3045,Webber:2003p3044}. As stressed above, for the sake of a meaningful comparison with data, it is important to take into account the stable isotopes of all elements. This is important not only for pure secondary elements, namely elements produced only through spallation, such as Boron (B=$^{10}$B+$^{11}$B), but also for Nitrogen (N=$^{14}$N+$^{15}$N), which gets a significant secondary contribution from spallation of heavier nuclei, for Carbon (C=$^{12}$C+$^{13}$C) and Oxygen (O=$^{16}$O+$^{17}$O+$^{18}$O) \cite[]{diBernardo:2010p10666,Evoli:2008p10682}. Moreover, for simplicity we assume an instantaneous decay for  isotopes whose lifetime is much shorter than their escape time from the Galaxy. This means that in the sum over $\alpha'$ in the rhs of Eq.~(\ref{eq:slab}) we consider also terms of the type $\sigma_{\alpha'\to\alpha''} f_{\alpha'}$ being $\alpha''$ a nuclear specie that rapidly decays into $\alpha$. 

Here $v(p)=\beta\, c$ is the velocity of nuclei of type $\alpha$ having momentum $p$. Notice that since the gas is assumed to be present only in the disc, and the ionization rate is proportional to the gas density, one can write: $\left(\frac{dp}{dt}\right)_{\alpha,ion}=2h_d \delta(z) b_{0,\alpha}(p)$, where $b_{0,\alpha}(p)$ contains the particle physics aspects of the process (see \cite{Strong:1998p10532} and references therein for a more detailed discussion of this term).  

Following the procedure outlined by \cite{aloisio1}, one can transform Eq.~(\ref{eq:slab}) in a modified weighted slab transport equation:
$$
\frac{I_{\alpha}(E)}{X_{\alpha}(E)} + \frac{d}{dE}\left\{\left[ \left(\frac{dE}{dx}\right)_{ad} +  \left(\frac{dE}{dx}\right)_{ion,\alpha}\right] I_{\alpha}(E)\right\} + \frac{\sigma_{\alpha} I_{\alpha}(E)}{m} =
$$
\be
~~~~~~~~~~~~= 2 h_d \frac{A_{\alpha} p^{2} q_{0,\alpha}(p)}{\mu\, v} + \sum_{\alpha'>\alpha} \frac{I_{\alpha}(E)}{m}\sigma_{\alpha'\to\alpha},
\label{eq:slab2}
\ee
where $I_{\alpha}(E)$ is the flux of nuclei with kinetic energy per nucleon E for nuclei of type $\alpha$, such that $I_{\alpha}(E)dE = v\,p^{2}f_{0,\alpha}(p)dp$. It is easy to show that $I_{\alpha}(E) = A_{\alpha}p^{2} f_{0,\alpha}(p)$, being $A_{\alpha}$ the atomic mass number of the nucleus.
We also introduced the quantity
\be
X_{\alpha}(E) = \frac{\mu \,v}{2 v_{A}} \left[ 1-\exp\left(-\frac{v_{A}}{D_{\alpha}}H\right)\right]
\ee
that represents the grammage for nuclei of type $\alpha$ with kinetic energy per nucleon $E$, while
\be
\left(\frac{dE}{dx}\right)_{ad} = -\frac{2 v_A}{3\mu\, c} \sqrt{E(E+m_p c^2)} 
\ee
is the rate of adiabatic energy losses due to advection. 

The diffusion coefficient relevant for a nucleus $\alpha$ can be written as:
\be
D_{\alpha} (p) = \frac{1}{3} \frac{p\,c}{Z_\alpha eB_{0}} v(p) \left[ \frac{1}{k\ W(k)} \right]_{k=Z_{\alpha} e B_{0}/pc},
\label{eq:diff}
\ee
where $W(k)$ is the power spectrum of waves at the resonant wavenumber $k=Z_{\alpha} e B_{0}/p\,c$. The non-linearity of the problem is evident here: the diffusion coefficient for each nuclear species depends on all other nuclei through the wave power $W(k)$, but the spectra are in turn determined by the relevant diffusion coefficient. The problem can be written in a closed form, though in an implicit way, by using the transport equation for each nucleus, Eq.~(\ref{eq:slab2}), and writing the evolution equation for the waves~\cite[]{Miller:1995p3113}:
\be
\frac{\partial}{\partial k}\left[ D_{kk} \frac{\partial W}{\partial k}\right] + \Gamma_{\rm CR}W = q_{W}(k), 
\label{eq:cascade}
\ee
where $q_{W}(k)$ is the injection term of waves with wavenumber $k$. In the present calculations we assume that waves are only injected on a scale $l_{c}\sim 50-100$ pc, for instance by supernova explosions. This means that $q_{W}(k)\propto \delta (k-1/l_{c})$. The level of pre-existing turbulence is normalized to the total power $\eta_B=\delta B^{2}/B_0^2 = \int dk W(k)$. Strictly speaking the wave number that appears in this formalism is the one in the direction parallel to that of the ordered magnetic field. In a more realistic situation in which most power is on large spatial scales, the role of the ordered field is probably played by the local magnetic field on the largest scale.

The term $\Gamma_{\rm CR}W$ in Eq. (\ref{eq:cascade}) describes the generation of wave power through CR induced streaming instability, with a growth rate \cite[]{Skilling:1975p2176}:
\be
\Gamma_{\rm cr}(k)=\frac{16 \pi^{2}}{3} \frac{v_{\rm A}}{k\,W(k) B_{0}^{2}} \sum_{\alpha} \left[ p^{4} v(p) \frac{\partial f_{\alpha}}{\partial z}\right]_{p=Z_{\alpha} e B_{0}/kc} ,
\label{eq:gammacr}
\ee 
where $\alpha$ is the index labeling nuclei of different types. All nuclei, including all stable isotopes for a given value of charge, are included in the calculations. As discussed in much previous literature, this is very important to compute properly the diffusion coefficient and thus for a meaningful comparison with  the flux spectra and  secondary to primary ratios, notably B/C. The growth rate, written as in Eq.~(\ref{eq:gammacr}), refers to waves with wave number $k$ along the ordered magnetic field. It is basically impossible to generalize the growth rate to a more realistic field geometry by operating in the context of quasi-linear theory, therefore we will use here this expression but keeping in mind its limitations.

The solution of Eq.~(\ref{eq:cascade}) can be written in an implicit form 

$$
W(k) = \left [ W_0^{1+\alpha_2}\left (\frac{k}{k_0}\right )^{1-\alpha_1} +\right .
$$
\be
~~~~~~~~~~~~\left . + \frac{1+\alpha_2}{C_{\rm K}v_A}\int_k^\infty\frac{dk'}{k'^{\alpha_2}}\int_{k_0}^{k'} d\tilde k \Gamma_{CR}(\tilde k) W(\tilde k) \right ]^{\frac{1}{1+\alpha_2}} ,
\label{eq:waves}
\ee
being $k_0=1/l_c$. In the present paper we assume a Kolmogorov phenomenology for the cascading turbulence, so that $\alpha_{1} = 7/2$ and $\alpha_{2} = 1/2$, and an unperturbed magnetic field $B_0 = 1 \mu G$. The two terms in Eq.~(\ref{eq:waves}) refer respectively to the pre-existing magnetic turbulence and the CR induced turbulence. In the limit in which there are no CRs (or CRs do not play an appreciable role) one finds the standard Kolmogorov wave spectrum 
\be
W(k)=W_0\left (\frac{k}{k_0} \right)^{-s} ~~~ s=\frac{\alpha_1-1}{\alpha_2+1} = \frac{5}{3}
\ee
normalized, as discussed above, to the total power $W_0=(s-1)l_c\eta_B$. 

The equations for the waves and for CR transport are solved together in an iterative way, so as to return the spectra of particles and the diffusion coefficient for each nuclear species and the associated grammage. The procedure is started by choosing guess injection factors for each type of nuclei, and a guess for the diffusion coefficient, which is assumed to coincide with the one predicted by quasi-linear theory in the presence of a background turbulence. The first iteration returns the spectra of each nuclear specie and a spectrum of waves, that can be used now to calculate the diffusion coefficient self-consistently. The procedure is repeated until convergence, which is typically reached in a few steps, and the resulting fluxes and ratios are compared with available data. This allows us to renormalize the injection rates and restart the whole procedure, which is repeated until a satisfactory fit is achieved. Since the fluxes of individual nuclei affect the grammage through the rate of excitation of streaming instability and viceversa the grammage affects the fluxes, the procedure is all but trivial. 

\section{Results}
\label{sec:results}

The main evidence for a transition from self-generated waves to pre-existing turbulence can be searched for in the spectra of the light elements, protons and helium nuclei. A spectral break was in fact found by the PAMELA experiment \cite[]{PAMELAbreak} in both spectra and later confirmed by AMS-02, although at the time of writing this paper only the results of AMS on protons have been published \cite[]{ams02}, while a preliminary version of the spectrum of helium has been presented \cite[]{amsslides}. The spectra of both elements were also measured by the Voyager \cite[]{voyager} outside the heliosphere, so as to make this the first measurement in human history of the CR spectra in the interstellar medium. This is a very important results in that it also allows us to refine our understanding of the effects of solar modulation \cite[]{potgieter2013}.

   \begin{figure}
   \centering 
   \includegraphics[width=8.cm]{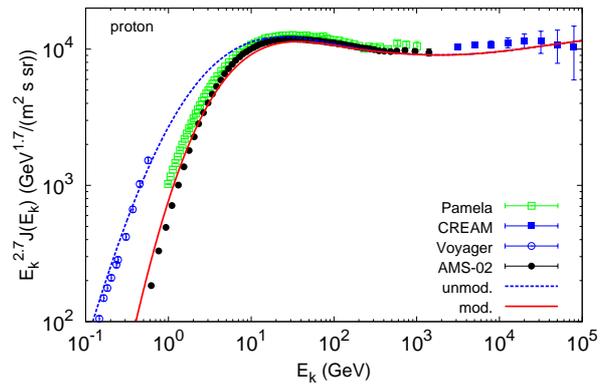}
   \caption{Spectrum of protons measured by Voyager (blue empty circles), AMS-02 (black filled circles) \cite[]{ams02}, PAMELA (green empty squares) \cite[]{PAMELAbreak} and CREAM (blue filled squares) \cite[]{cream}, compared with the prediction of our calculations (lines). The solid line is the flux at the Earth after the correction due to solar modulation, while the dashed line is the spectrum in the ISM. %At low energies the predicted flux is in excellent agreement with the Voyager data. "Io evito commenti di merito nelle caption!"
}
              \label{fig:spectrap}%
    \end{figure}

   \begin{figure}
   \centering 
   \includegraphics[width=8.cm]{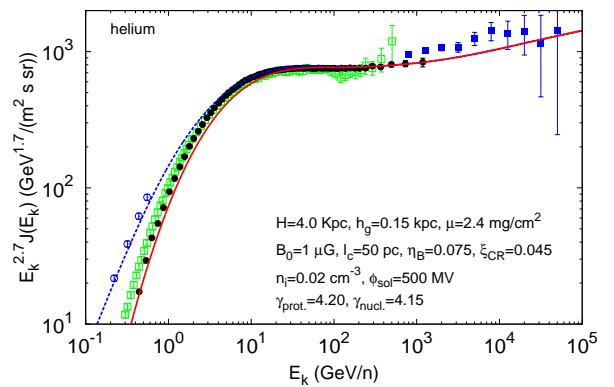}
   \caption{Spectrum of He nuclei according to preliminary measurements of AMS-02 (black filled circles), as measured by Voyager (blue empty circles), PAMELA (green empty squares) and CREAM (blue filled squares), compared with the prediction of our calculations (lines). The solid line is the flux at the Earth after the correction due to solar modulation, while the dashed line is the spectrum in the ISM. %At low energies the predicted flux is in excellent agreement with the Voyager data. "Io evito commenti di merito nelle caption!"
   }
              \label{fig:spectraHe}%
    \end{figure}

The spectrum of protons and helium nuclei as calculated in this paper is shown in Figs. \ref{fig:spectrap} and \ref{fig:spectraHe}, respectively: the solid lines indicate the spectra at the Earth, namely after solar modulation modelled using the force-free approximation \cite[]{gleeson68}, while the dashed lines are the spectra in the ISM. The data points are the spectra measured by the Voyager (empty circles) \cite[]{voyager}, AMS-02 (filled circles) \cite[]{ams02}, PAMELA (empty squares) \cite[]{PAMELAbreak} and CREAM (filled squares) \cite[]{cream}. Figs. \ref{fig:spectrap} and \ref{fig:spectraHe} show several interesting aspects: 1) both the spectra of protons and helium nuclei show a pronounced change of slope at few hundred GeV/n, where self-generation of waves becomes less important than pre-existing turbulence (in fact, the change of slope takes place in rigidity). 2) We confirm that injecting He with a slightly harder spectrum with respect to protons ($p^{-4.15}$ versus $p^{-4.2}$) improves the fit to the data. 3) The spectra calculated to optimise the fit to the AMS-02 and PAMELA data is in excellent agreement with the Voyager data (see dashed lines). This is all but trivial: in our model, at sufficiently low energies (below $\sim 10$ GeV/n), particle transport is dominated by advection (at the Alf\'en speed) with self-generated waves rather than diffusion. This reflects into a weak energy dependence of the propagated spectra that is exactly what Voyager measured (see also \cite[]{potgieter2013}). 4) At low energies, the agreement of the predicted spectra with those measured by Voyager is actually better than the agreement with the modulated spectra, as observed with AMS--02; this suggests that probably the prescriptions used to describe solar modulation are somewhat oversimplified, either when applied to data collected over extended periods of time, when the effective solar potential may change appreciably, or because of intrinsic limitations of the force-field approximation.

\begin{figure}
   \centering 
   \includegraphics[width=8.cm]{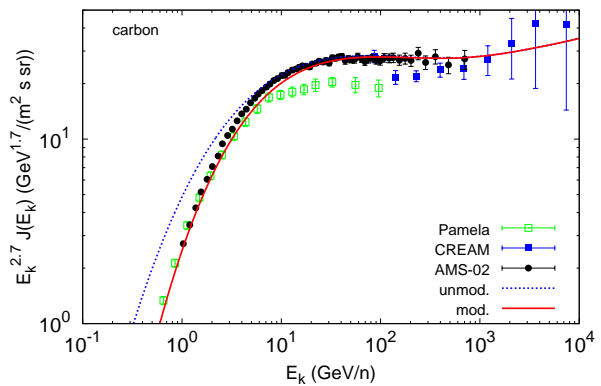}
   \caption{Spectrum of C nuclei as measured by CREAM (blue squares), PAMELA (green empty squares), and according to preliminary measurements of AMS-02 (black circles), compared with the prediction of our calculations (lines). The solid line is the flux at the Earth after the correction due to solar modulation, while the dashed line is the spectrum in the ISM.}
              \label{fig:carbon}%
    \end{figure}

For each heavier nucleus, we assume the same {\it injected spectral shape} in rigidity as for helium, keeping as only free parameter the normalization, chosen to match the data. In Fig.~\ref{fig:carbon} we illustrate the prediction for Carbon nuclei (which is also a needed ingredient to compute the B/C ratio), compared with data by PAMELA and CREAM, as well as preliminary data by AMS-02. 
The free normalization is chosen to match more closely the AMS-02 data.  Clearly, the phenomenon of transition from self-generated to pre-existing waves manifests itself in the transport of all nuclei, hence we should expect a spectral break at the same rigidity as for helium and protons. This prediction appears currently in agreement with Carbon spectrum observations, although it is hard to judge to what extent a break is present in AMS-02 data alone, giving the growing error bars and the limited dynamical range at high energy. Note that a break would appear more prominent if one were to combine PAMELA and CREAM data,
which do seem to differ from AMS-02 data in the 10 to $\sim 200$ GeV/n range beyond the reported errors. Definitely, the forthcoming AMS-02 publication of nuclear fluxes should
help in clarifying the situation.

\begin{figure}
   \centering 
   \includegraphics[width=8.cm]{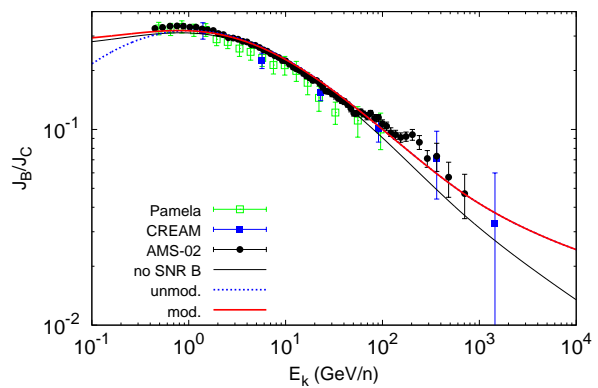}

   \caption{B/C ratio  as measured by CREAM (blue squares), PAMELA (green empty squares), and according to preliminary measurements of AMS-02 (black circles). The black/bottom solid line is the prediction of our model, while the red/top line has been obtained by adding a source grammage of $0.15\,$g\,cm$^{-2}$, close to that given by Eq.~(\ref{eq:source}).}
              \label{fig:sec}%
    \end{figure}
    
    In Fig. \ref{fig:sec} we show the calculated B/C ratio (solid black line) as compared with data from CREAM (blue squares), PAMELA (green squares), and the still preliminary ones from AMS-02 (black circles).  Even if the injected Carbon flux is normalized to the preliminary Carbon data reported by AMS-02, the B/C ratio is still in satisfactory agreement
with  both PAMELA and CREAM data, as for our previous result \cite[]{aloisio1}.
The B/C ratio  also fits the AMS data up to $\sim 100\,$GeV/n. At higher energy, the AMS-02 analysis seems to suggest a B/C ratio somewhat higher than our prediction. While its significance is uncertain, given the preliminary nature of AMS data, if this ``excess'' is interpreted  as physical, it would suggest the presence of an additional contribution to the grammage traversed by CRs. The most straightforward possibility to account for such a grammage is that it may be due to the matter traversed by CRs while escaping the source, for instance a SNR. The grammage due to confinement inside a SNR can be easily estimated as 
\be
X_{\rm SNR} \approx 1.4 r_{s} m_{p} n_{\rm ISM} c T_{\rm SNR} \approx 0.17\, {\rm g\,cm^{-2}} \frac{n_{\rm ISM}}{{\rm cm}^{-3}} \frac{T_{\rm SNR}}{2\times 10^{4}{\rm yr}},
\label{eq:source}
\ee
where $n_{\rm ISM}$ is the density of the interstellar gas upstream of a SNR shock and $r_{s}=4$ is the compression factor at the shock and $T_{\rm SNR}$ is the duration of the SNR event (or better, the lifetime ``useful'' to confine particles up to $E\sim$TeV/n), assumed here to be of order twenty thousand years. The factor 1.4 in Eq.~(\ref{eq:source}) has been introduced to account for the presence of elements heavier than hydrogen in the target. While Eq.~(\ref{eq:source}) is only a rough estimate of the grammage at the source, in that several (in general energy dependent) factors may affect such an estimate, at least it provides us with a reasonable benchmark value. The solid red curve in Fig.~\ref{fig:sec} shows the result of adding the grammage accumulated by CRs inside the source to the one due to propagation in the Galaxy. It is clear that by eye it fits better the AMS-02 data at high rigidity, while being also compatible with the older CREAM data. The forthcoming publication by AMS-02 of the fluxes of nuclei and secondary to primary ratios should hopefully clarify the situation and provide more reliable discriminating power. 

%\bibliographystyle{aa}
%\bibliography{SelfGenBib} 

\section{Summary and Conclusions}
\label{sec:summary}

One of the main implications of the transport theory of CRs in the Galaxy is that the spatial gradient developed due to the diffusive propagation of CRs (for instance tuned to fit the B/C ratio) is such that the rate of growth of Alfv\'en waves cannot be neglected, and in fact at rigidity $\lesssim 200$ GV the self-generated waves must play a dominant role. On the other hand, at higher rigidity the growth rate drops with respect to damping processes responsible for cascading of waves in wavenumber space. As shown by \cite{serpico1} and later by \cite{aloisio1} the transition between the two reflects in a change of slope in the spectra of primary elements, most notably protons and helium nuclei. Evidence in favour of such a break emerged first in nuclear spectra~\cite[]{Ahn:2010gv}; then  hints for something similar in proton and helium spectra were suggested by the need to reconcile high-energy data measured by CREAM~\cite[]{cream} with low-energy ones; more recently, PAMELA \cite[]{PAMELAbreak} reported for the first time the
detection of the transition in proton and helium nuclei; finally, after contradictory early results presented by AMS-02 in 2013~\cite[]{AMS13p,AMS13He},  this behaviour for proton and helium has been confirmed by the AMS-02 collaboration \cite[]{ams02,amsslides}, albeit the publication only includes the proton spectrum, with the helium flux being still preliminary.

The results found in this paper can be summarised as follows: 1) both the proton and helium spectra show a pronounced change of slope at rigidity of a few hundred GV, which we interpret as the region where the self-generation of waves becomes less important than pre-existing turbulence. Our interpretation implies that a similar spectral break should also be present in the spectra of other primary nuclei (for instance carbon); current data are consistent with this prediction, although the uncertainties, both the ones of statistical nature and possibly others of systematic origin, do not enable a stringent quantitative test, so that no final conclusion on this issue can be drawn, yet. 2) The observed difference in spectral slope between He nuclei and protons should be attributed to differences at acceleration/injection, with the injection spectrum of He nuclei that must be somewhat harder ($\propto p^{-4.15}$) compared with the proton one ($\propto p^{-4.2}$). A consensus on the theoretical explanation of this difference is not yet available, although some models have been put forward, for instance involving: different acceleration sites (as in the reverse shock of~\cite{Ptuskin:2012qu},
%~\footnote{Anche se penso che questo non funzioni!})
the inhomogeneous chemical composition of the medium at the acceleration site~\cite[]{OI2011}, or phenomena associated to the physics of the injection process~\cite[]{malkov}, such  as that the more easily ionized hydrogen gets a greater contribution by  older and weaker shocks (associated to steeper spectra)~\cite[]{Drury:2010am}.
3) The spectra of protons and He nuclei predicted by our model automatically fit the low energy Voyager data, which is remarkable, because this suggests that the propagation of low rigidity ($\lesssim 10$ GV) protons and He nuclei is dominated by advection with self-generated Alfv\'en waves, as was originally proposed by \cite{serpico1}. 4) Data on secondaries or secondary to primary ratios pre-AMS are consistent with the picture above, as already shown in~\cite[]{aloisio1}, but do not allow for a stringent test of this scenario, yet. Preliminary AMS-02 data on the B/C ratio show greater potential for discriminating power and, if taken at face value, provide marginal evidence for an excess grammage at energy/nucleon  $\gtrsim 100$ GeV/n. We speculated that this grammage may be the one traversed by CRs in the acceleration region: in fact, a quick  estimate of the equivalent grammage in a typical SNR (putative source of CRs) compares exceptionally well with the grammage needed to reconcile the predicted and the observed one. 
A generic expectation is that  at Energies/nucleon of {\cal O}(TeV/n), the contribution of production at the ``sources'' for so-called secondary species may become significant,
as already noted in the past for other species such as antiprotons~\cite[]{Blasi:2009bd}. Needless to say, current B/C data presented by the AMS-02 collaboration are still preliminary and
one should not over-interpret them, especially in the light of the challenging path that lead to the final establishment of the presence of changes of slope in the proton (and most likely also helium) spectrum. Yet, it is certainly true that secondary-to-primary ratios, at the level of precision and dynamical range that should be achievable by AMS-02, carry a substantial amount of information, as illustrated by our comparison. A more detailed re-analysis of our model will be certainly justified, once AMS-02 data on nuclear species (both primaries and secondary ones) will be eventually  published. 

\section*{Acknowledgements}
PB is grateful to the members of the Arcetri High Energy Astrophysics Group for continuous insightful conversations. The work of PB was partially funded through Grant PRIN-INAF 2012. PS thanks the Arcetri Astrophysical Observatory, where part of this work was done, for the very kind hospitality.

\end{document}